# On the Study of Chaos and Memory Effects in the Bonhoeffer-van der Pol Oscillator with a Non-Ideal Capacitor

Jamieson Brechtl[a], Xie Xie[b], Karin A. Dahmen[c], Peter K. Liaw[b], and Steven J. Zinkle[b,d]

a: The Bredesen Center for Interdisiplinary Research and Education, The University of Tennessee, Knoxville TN 37996, USA

b: Department of Materials Science and Engineering, The University of Tennessee, Knoxville, TN 37996, USA

c: Department of Physics, The University of Illinois at Urbana-Champaign, Champaign, IL 61820, USA

d: Department of Nuclear Engineering, The University of Tennessee, Knoxville, TN 37996, USA

*Abstract*

In this paper, the voltage fluctuations of the Bonhoeffer van der pol oscillator system with a non-ideal capacitor were investigated. Here, the capacitor was modeled, using a fractional differential equation in which the order of the fractional derivative, $α$, is also a measure of the memory in the dielectric. The governing fractional differential equation was derived using two methods, namely a differential and integral approach. The former method utilized a hierarchical resistor-capacitor (RC) ladder model while the latter utilized the theory of the universal dielectric-response. It was found that the dynamical behavior of the potential across the capacitor was affected by this parameter, and, therefore, the memory of the system. Additionally, findings indicate that an increase in the memory parameter was associated with an increase in the energy stored in the dielectric. Furthermore, the effects of the dynamical behavior of the voltage on the capacity of the dielectric to store energy was examined. It was found that oscillation death resulted in a higher amount of stored energy in the dielectric over time, as compared to behavior, which displayed relaxation oscillations or chaotic fluctuations. The relatively-lower stored energy resulting from the latter types of dynamical behavior appeared to be a consequence of the memory effect, where present accumulations of energy in the capacitor are affected by previous decreases in the potential. Hence, in this type of scenario, the dielectric material can be thought of as "remembering" the past behavior of the voltage, which leads to either a decrease, or an enhancement in the stored energy. Moreover, an increase in the fractional parameter $\alpha$, under certain conditions, led to the earlier onset of the chaotic voltage oscillations across the capacitor. Furthermore, the corresponding phase portraits showed that the chaotic behavior was heightened, in general, with a decrease in $\alpha$. The non-ideal capacitor was also found to have a transitory nature, where it becomes more like a resistor as $\alpha \to 0$, and conversely, more like a capacitor as $\alpha \to 1$. Here, a decrease in $\alpha$ was linked to an enhanced metallic character of the dielectric. Finally, a possible link between the complexity of the voltage noise fluctuations and the metallic character of the non-ideal capacitor will be discussed.

## 1. Introduction

Nonlinear systems are a subject of great importance since they have a wide range of applications in many fields, including physics, nonlinear mechanics, and optics [1-4]. One such tool, which is used to study such phenomenon, includes the Van der Pol (VP) equation, which was introduced by B. Van der Pol in his 1920 paper [5]. This type of system can exhibit the chaotic behavior that is characterized by the aperiodic behavior, which depends sensitively on the initial conditions [6]. Importantly, this system has played a critical role in the development of nonlinear dynamics.

The Bonhoeffer–Van der Pol (BVP) oscillator [7-10] is a modified version of the VP oscillator in which the underlying dynamics of the BVP oscillator are equivalent to a simplified Hodgkin–Huxley model [11, 12]. This type of scheme has also been considered as a prototype model for excitable systems. Given the right conditions, the BVP can generate th complicated mixed-mode oscillations (MMOs), which are characterized by alternating large- and small-amplitude excursions in the observed time series [14, 15]. Besides the BVP system, MMOs have been observed in chemical and neuronal firing systems [16-19].

As will be discussed later, a major circuit element of the BVP oscillator is a capacitor. In the past, capacitor models that involve a 'universal' dielectric response have been proposed [20, 21]. The model was originally based on Curie's empirical law [22], which states that the current produced in a dielectric material exhibits the time-dependent power-law behavior that is proportional to the applied voltage, which is constant. However, Westerlund et al. expanded upon Curie's model to include a general input voltage $v$(t) [20]. At the time, their model contained properties that others did not have, such as the dielectric absorption and insulation resistance.

Interestingly, the capacitor model discussed in [20] uses the fractional calculus, which involves derivatives and integrals of arbitrary order [23]. Interestingly, these types of operators involve the integration of the original function multiplied by a power-law function [24]. According to Tarasov, this kind of mathematical operation is also a type of power-like memory function [25]. In the present context, memory means that the present state evolution of a system depends on all past states. Furthermore, long-term memory effects correspond to intrinsic dissipative processes in different physical systems [26, 27].

This type of formalism is what led Westerlund to conclude that a non-ideal dielectric capacitor will have memory of all earlier voltages. Applying this type of operation to a dielectric capacitor led Westerlund to conclude that these types of materials possess memory effects. Here he suggested that these memory effects are closely related to the dielectric absorption that occurs during charging [28].

In a circuit whose governing equations consist of non-integer order derivatives (and integrals), the current across the capacitor is proportional to the fractional derivative of the applied potential [29, 30]. In terms of the capacitor model, values of $\alpha$, which denote the order of the derivative, may depend on the type of the dielectric material [20]. For example, in [20, 21, 31], it was reported that $\alpha$ ranged between $0.6 - 1$, where $\alpha = 0.6$ for $SiO_x$, while $\alpha = \sim 1$ for polypropylene.

Besides circuit models, this branch of mathematics has been found to be useful in modeling and analyzing phenomena, such as rheology [32-36], quantum physics [25, 37, 38], and diffusion [39-41]. Additionally, the non-integer order calculus has other advantages, such as the ability to model complex systems that contain long-range spatial interactions and memory effects [42].

Moreover, the use of the fractional calculus is convenient in modeling circuits since the equations used closely resemble the traditional equations already used.

A major goal of the present work, therefore, is to examine how inducing memory effects alters the underlying dynamics of the BVP oscillator. Inducing memory into the circuit will be accomplished by exchanging a typical capacitor in the circuit model with a non-ideal one. Furthermore, it is also the interest of the current work to examine the interplay between the memory effects and the chaotic behavior of the system. It is believed that this innovative approach will advance our fundamental knowledge on the effects of chaos and memory in the Bonhoeffer van der Pol oscillator system.

## 2. Modeling and Analysis

Figure 1 shows the circuit diagram of the forced BVP oscillator [43]. The electric circuit consists of an inductor, *L*, a capacitor, *C*, a linear resistor, *R*, a nonlinear negative conductance, *G*, and two voltage sources, $V_0$, and $V_1 \sin\omega t$. The nonlinear current, $g(v_c)$, will be expressed as the following cubic polynomial [44]:

$$g(v_c) = \beta v_c^3 - \gamma v_c \; ; \quad \beta > 0 \; \gamma > 0 \tag{1}$$

The governing equations of the electric circuit, based on Kirchhoff's law, may be represented by the following equations:

$$i_c = C \frac{dv_c}{dt} = i - g(v_c) \tag{2}$$

$$L \frac{di}{dt} = -v_c - iR + V_0 + V_1 \sin\omega t \tag{3}$$

As was done in [43], we normalize Equations (2) and (3) by setting:

$$C = \varepsilon\gamma^2 L,\; R = \frac{k_1}{\gamma},\; V_0 = \sqrt{\frac{\gamma}{\beta}} B_0,\; V_1 = \sqrt{\frac{\gamma}{\beta}} B_1,\; v_c = \sqrt{\frac{\gamma}{\beta}} \bar{v}_c,\; i = \sqrt{\frac{\gamma^3}{\beta}} \bar{\imath},\; t = \gamma L \xi,\; \text{and}\; \frac{d}{d\xi} = \cdot \tag{4}$$

Substituting $\frac{d\tilde{v}_c}{d\xi}$ in for the normalized capacitor current and expanding $g(v_c)$ give:

$$\frac{d\bar{v}_c}{d\xi} = \frac{1}{\varepsilon}\left[\bar{\imath} - (\bar{v}_c^3 - \bar{v}_c)\right] \tag{5}$$

$$\frac{d\bar{\imath}}{d\xi} = -\bar{v}_c - k_1 \bar{\imath} + B_0 + B_1 \sin\bar{\omega}\xi \tag{6}$$

As was done in [43], the parameters, $\varepsilon$, $k_1$, and $\bar{\omega}$, were set to 0.1, 0.9, and 1.35 respectively. From here the non-ideal capacitor, in which the current is equal to the fractional order of the applied potential, will be derived in the next couple of sections, using both an integral and a differential method.

## 2.1 Integral Method

In terms of the capacitor, models have been proposed, which considers a 'universal' response of the dielectric material [21]. More specifically, this response is related to the ability of the dielectric system to retain the 'memory' of past excitations. Like the viscoelastic damper, the constitutive relationship between the polarization of the dielectric and a time-varying electric field can be written in terms of the convolution of a response function with the electric field:

$$P(t) = \varepsilon_0 \int_0^\infty f(u)E(t-u)du \quad (7)$$

where $P(t)$ is the polarization of the dielectric, $\varepsilon_0$ is the dielectric permittivity of the free space, $f(u)$ is the dielectric response function, and $E(t)$ is the electric field between the plates of the capacitor. In addition, taking the Fourier transform of $f(u)$ gives the frequency-dependent susceptibility, $\chi(\omega) = \chi'(\omega) - i\chi''(\omega)$, of the medium. Taking the derivative of the polarization with respect to time yields the current flowing through the capacitor:

$$i_c(t) = \frac{dP}{dt} = \varepsilon_0 \frac{d}{dt}\int_0^\infty f(u)E(t-u)du = \varepsilon_0 \frac{d}{dt}\int_{-\infty}^t f(t-\tau)E(\tau)d\tau \quad (8)$$

where the substitution, $\tau = t - u$, was used. Next, we assume that the response function is composed of a power-law function, which is also related to the susceptibility of the dielectric medium. To satisfy this condition, we set the dielectric response function, $f(t-\tau) = \frac{A}{(t-\tau)^\alpha}$ ($0 < \alpha < 1$), where $A$ is some constant of proportionality:

$$i_c(t) = A\varepsilon_0 \frac{d}{dt}\int_{-\infty}^t \frac{E(\tau)}{(t-\tau)^\alpha}d\tau = A\varepsilon_0 \frac{d}{dt}\int_{-\infty}^t \frac{E(\tau)}{(t-\tau)^\alpha}d\tau \quad (9)$$

It should be stated that the power-law behavior of *f(t)* is based on the Curie-von Schweidler law, which is universal [45]. For frequencies in the sub-audio to ~$10^9$ Hz range, the corresponding dielectric loss term is proportional to $\omega^{\alpha-1}$ [46]. Due to its ubiquitous nature, the law can be applied to all types of chemical bonds, all possible types of polarizing species, and to materials containing single crystals, polycrystalline, and amorphous structures [21].

Now let us assume that at t < 0, there is no current flowing through the system and, thus, there is a negligible electric field emanating from the capacitor plate. Therefore, we may change the lower limit of the integral in Equation (9) to zero. Moreover, assuming the spatial uniformity of the electric field along the distance perpendicular to the capacitor plates allows one to write $E(t) = \frac{v(t)}{\Delta z}$, where $\Delta z$ is the distance between the opposing sides:

$$i_c(t) = A \frac{\varepsilon_0}{\Delta z} \frac{d}{dt} \int_0^t \frac{v(\tau)}{(t-\tau)^\alpha} d\tau = \frac{\Gamma(1-\alpha)}{h_1} {}_0 D_t^\alpha v(t) \quad (10)$$

where $h_1$ equals $\frac{\Delta z}{A\varepsilon_0}$, and has units of $\Omega s^{-\alpha}$ [28]. Before moving on, it should be mentioned that an integral similar to the one in Equation (10) was derived by von-Schweidler [28, 47]. Setting $C_\alpha = \frac{\Gamma(1-\alpha)}{h_1}$ and simplification give:

$$i_c(t) = C_\alpha {}_0 D_t^\alpha v(t) \quad (11)$$

It should be noted that the equation for the coefficient, $C_\alpha$, above and Equation (11) are the same as for the capacitor model proposed by Westerlund et al. [20, 28].

## 2.2 Differential Method

In the past, the RC Ladder models have been used to describe the time-dependent current profile in a circuit system [23,48, 49]. However, for the present work, a hierarchical model, as proposed by Schiessel et al., will be used [32-34, 50]. This type of model is characterized by a tiered structure whose behavior is described by continued fraction expressions. A similar mathematical framework as the one shown below was used to describe a finite n-section-lumped RC ladder system [51]. Figure 2 presents the arrangement of the model, which consists of resistor elements on the strut and a capacitor element beneath the resistor. For the modeling and analysis, the following conditions will be assumed:

$$i(t \leq 0) = i_j(t \leq 0) = v(t \leq 0) = v_j(t \leq 0) = 0 \quad ; \quad j \in [0, n] \tag{12}$$

Applying Ohm's law to the resistor, $R_0$, and capacitor, $C_0$, of the circuit model from Figure 2 yields:

$$v(t) = R_0 i(t) + v_0(t) \quad ; \quad i(t) = i_1(t) + C_0 \frac{d}{dt} v_0(t) \tag{13}$$

Algebraically manipulating Equation 13 after applying the Laplace transform and initial condition gives:

$$\frac{\tilde{v}(s)}{R_0 \tilde{i}(s)} = 1 + \frac{\frac{1}{R_0 C_0}}{s + \frac{\tilde{i}_1(s)}{C_0 \tilde{v}_0(s)}} \tag{14}$$

To generalize Equation (14), we apply Ohm's law to the kth element in the circuit:

$$v_k(t) = R_{k+1} i_{k+1}(t) + v_{k+1}(t) \quad ; \quad i_k(t) = i_{k+1}(t) + C_k \frac{d}{dt} v_k(t) \tag{15}$$

Applying the Laplace transform to Equation (15), and using the above method, we can get the following iterative formula:

$$\frac{1}{R_{k+1}} \frac{\tilde{v}_k(s)}{\tilde{\imath}_{k+1}(s)} = 1 + \frac{\frac{1}{R_{k+1}C_{k+1}}}{s + \frac{\tilde{\imath}_{k+2}(s)}{C_{k+1}\tilde{v}_{k+1}(s)}} \quad ; \quad k \in [0, n-2] \tag{16}$$

To find the nth term for the ladder model, we plug in $k = n - 2$ into Equation (16) with the condition, $\tilde{v}_{n-1}(s) = \tilde{v}_n(s) = R_n\tilde{\imath}_n(s)$ (based on Figure 2), which yields:

$$\frac{1}{R_{n-1}} \frac{\tilde{v}_{n-2}(s)}{\tilde{\imath}_{n-1}(s)} = 1 + \frac{\frac{1}{R_{n-1}C_{n-1}}}{s + \frac{1}{R_n C_{n-1}}} \tag{17}$$

Therefore, combining Equations (15) - (17), and expanding upon them, gives the following continued fraction form:

$$\frac{\tilde{v}(s)}{R_0 \tilde{\imath}(s)} = 1 + \frac{s^{-1}\left(\frac{1}{R_0 C_0}\right)}{1+} \frac{s^{-1}\left(\frac{1}{R_1 C_0}\right)}{1+} \frac{s^{-1}\left(\frac{1}{R_1 C_1}\right)}{1+} \frac{s^{-1}\left(\frac{1}{R_2 C_1}\right)}{1+} \cdots \frac{s^{-1}\left(\frac{1}{R_{n-1}C_{n-1}}\right)}{1+} \frac{s^{-1}\left(\frac{1}{R_n C_{n-1}}\right)}{1} \tag{18}$$

For reasons that will be made obvious, we introduce the continued fraction form for $y(1 + y)^{\alpha-1}$ via Equation (11.7.2) from [52]:

$$y(1+y)^{\alpha-1} = \lim_{n \to \infty} \frac{y}{1+} \frac{(1-\alpha)y}{1+} \underset{m=3}{\overset{n}{\mathbf{K}}} \frac{a_m y}{1+} \tag{19}$$

$$a_{2i} = \frac{i - \alpha}{2(2i - 1)} \qquad a_{2i+1} = \frac{i + \alpha - 1}{2(2i - 1)} \tag{20}$$

where y = $s^{-1}\left(\frac{1}{R_0 C_0}\right)$. Expanding out the series from Equation (18) for $i \in [3, n-2]$ gives:

$$y(1+y)^{\alpha-1} \approx \frac{y}{1+} \frac{(1-\alpha)y}{1+} \frac{1 \cdot (0+\alpha)}{1 \cdot 2} y \frac{1 \cdot (2-\alpha)}{2 \cdot 3} y \frac{2 \cdot (1+\alpha)}{3 \cdot 4} y \frac{n-2-\alpha}{2 \cdot (2n-5)} y \quad (21)$$

Setting $\lambda = \frac{1}{R_0 C_0}$ and comparing term-by-term Equations (18) and (21) yields:

$$\frac{1}{R_1 C_0} = (1-\alpha)\lambda \;\; ; \;\; \frac{1}{R_1 C_1} = \frac{\alpha}{2}\lambda \;\; ; \cdot \frac{1}{R_{n-1} C_{n-1}} = \frac{n+\alpha-2}{2(2n-3)}\lambda \;\; ; \;\; \frac{1}{R_n C_{n-1}} = \frac{n-\alpha}{2(2n-1)}\lambda \quad (22)$$

From the above equation, each $C_k$ and $R_k$ may be written in terms of their respective $C_0$, $R_0$ as [32]:

$$R_k = (2k-1) \frac{\Gamma(1-\alpha)}{\Gamma(\alpha)} \frac{\Gamma(k+\alpha-1)}{\Gamma(k-\alpha+1)} R_0 \quad (23)$$

$$C_k = 2 \frac{\Gamma(\alpha)}{\Gamma(1-\alpha)} \frac{\Gamma(k-\alpha+1)}{\Gamma(k+\alpha)} C_0 \quad (24)$$

If we substitute the terms from Equation (22) into Equation (18) and take the limit $n \to \infty$, we obtain:

$$\frac{\tilde{v}(s)}{R_0 \tilde{\imath}(s)} = 1 + \left(\frac{\lambda}{s}\right)\left(1 + \frac{\lambda}{s}\right)^{\alpha-1} \quad (25)$$

Now, Equation (25) may be converted into the following form:

$$\frac{\tilde{v}(s)}{R_0 \tilde{\imath}(s)} = \left(\frac{\lambda}{s}\right)^{\alpha} = \left(\frac{1}{R_0 C_0 s}\right)^{\alpha} \quad (26)$$

, which is valid for $6R_0C_0n^2 \leq 1/s \leq \frac{1}{6}R_0C_0n^2$ [23]. Now, taking the inverse Laplace transform of Equation (26) in this range yields:

$$i(t) = R_0^{\alpha-1}C_0^{\alpha}D_t^{\alpha}v(t) \quad ; \quad 6R_0C_0 \leq t \leq \frac{1}{6}R_0C_0n^2 \tag{27}$$

It should also be stated that given enough elements in the capacitor, the relation from the above equation is preserved for long times. However, if we are dealing with a dielectric material that exhibits a loss peak in certain frequency ranges, a further restriction must be applied:

$$6R_0C_0 \leq t < \frac{1}{\omega_p} \tag{28}$$

where $\omega_p$ is the frequency at which the peak occurs. Therefore, if a loss peak is observed, the condition imposed by Equation (28) is necessary for the application of the Curie-von Schweidler law [i.e., $f(t) \propto t^{\alpha}$] to Equation (9). To increase the reduced time range imposed by the large peak frequencies, one can select sufficiently small-valued resistors and capacitors. As discussed in [32], $R_k$ and $C_k$ may be chosen in such a way that $R_0 = R$ and $C_0 = C$. In doing so, we obtain:

$$i_c(t) = R^{\alpha-1}C^{\alpha}{}_0D_t^{\alpha}v_c(t) \tag{29}$$

Notice the similarities between the above equation and Equation (11). In fact, they only differ in their coefficients, where the former has a coefficient defined as $C_{\alpha}$, while the latter is given by $R^{\alpha-1}C^{\alpha}$. It should also be stated that given enough elements in the capacitor, the relation from Equation (27) is preserved. As observed in Equation (29), when $\alpha = 1$, the circuit element behaves as a standard capacitor, while for $\alpha = 0$, the device behaves purely as a resistor [20, 28]. Therefore, a dielectric capacitor in this context can be thought of as one that behaves in a fashion that is intermediate between those of a standard capacitor and a resistor.

In terms of the behavior of the above model, the following may explain its underlying dynamics. As a voltage, $\theta_0 v(t)$ ($\theta_0$ is the Heaviside step function), is suddenly applied to the capacitor system (Figure 2), the potential is only experienced by the first resistor, $R_0$. Here, the element, $R_0$, in addition to the other $R_i$'s, may represent some resistive or energy-loss mechanism. After a brief period, the voltage on the first capacitor, $C_0$, begins to rise, which represents the storage of energy in the structure. Simultaneously, the current flows across the next resistor, $R_1$, creating a potential across the element. In this way, the voltage propagates continuously down the ladder, inducing the energy transfer along the way.

Almeida et al. put forth the following approximation for the fractional derivative [53]:

$$_aD_t^\alpha x(t) \approx$$
$$A(N,\alpha)(t-a)^{-\alpha}x(t) + B(N,\alpha)(t-a)^{1-\alpha}\dot{x}(t) - \sum_{p=2}^{N} C(p,\alpha)(t-a)^{1-p-\alpha}V_p(t) = \phi(t,x,\dot{x}) \quad (30)$$

where $0 < \alpha < 1$. Additionally, $A(N, \alpha)$, $B(N, \alpha)$, $C(p, \alpha)$, and $V_p(t)$ are defined as the following:

$$A(N,\alpha) = \frac{1}{\Gamma(1-\alpha)}\left[1 + \sum_{p=2}^{N} \frac{\Gamma(p-1+\alpha)}{\Gamma(\alpha)(p-1)!}\right] \quad (31)$$

$$B(N,\alpha) = \frac{1}{\Gamma(2-\alpha)}\left[1 + \sum_{p=1}^{N} \frac{\Gamma(p-1+\alpha)}{\Gamma(\alpha-1)p!}\right] \quad (32)$$

$$C(p,\alpha) = \frac{1}{\Gamma(2-\alpha)\Gamma(\alpha-1)} \frac{\Gamma(p-1+\alpha)}{(p-1)!} \quad (33)$$

$$V_p(t) = (1-p)\int_a^t (\tau-a)^{p-2}x(\tau)d\tau \quad ; \quad p \in [2,N] \quad p \in \mathbb{N} \quad (34)$$

The moments, $V_p$, are regarded as solutions to the following system of differential equations:

$$\dot{V}_p(t) = (1-p)(t-a)^{p-2}x(t) \tag{35}$$

$$V_p(a) = 0, \forall p \in [2, N] \tag{36}$$

The power of Equations (30) - (36) lies in their ability to approximate the fractional derivative, using the original function and its first order derivative. In addition, the above method only requires ~ 10 terms for a relatively-good approximation of the fractional derivative, whereas methods, such as Grünwald's definition [54], require significantly more terms. However, Equation (30) will produce singularities and complex values when $t \leq a$, which is an issue since $t \geq 0$. To avoid this problem, $a$ was chosen such that it is negative and sufficiently close to zero.

Now, plugging in $\frac{d^\alpha}{d\xi^\alpha}\bar{v}_c(\xi)$ for the normalized capacitor current in Equation (5) gives:

$$\frac{d^\alpha \bar{v}_c}{d\xi^\alpha} = \frac{1}{\varepsilon}\left[\bar{\imath} - \left(\bar{v}_c^3 - \bar{v}_c\right)\right] \tag{37}$$

$$\frac{d\bar{\imath}}{d\xi} = -\bar{v}_c - k_1\bar{\imath} + B_0 + B_1 sin\bar{\omega}\xi \tag{38}$$

Using Equation (30), we may approximate $\frac{d\bar{v}_c}{d\xi}$, which yields:

$$\frac{d\bar{v}_c}{d\xi} \approx \varphi(\xi)\left\{\frac{1}{\varepsilon}[\bar{\imath}(\xi) - \bar{v}_c^3(\xi) + m_1(\xi)\bar{v}_c(\xi)] + m_2(\xi)\right\} \tag{39}$$

$$m_1(\xi) = 1 - A(\alpha, N)(\xi - a)^{-\alpha} \tag{40}$$

$$m_2(\xi) = \sum_{p=2}^{N} C(p, \alpha)(\xi - a)^{1-p-\alpha} V_p(\xi) \tag{41}$$

$$\varphi(\xi) = \frac{1}{B(\alpha, N)}(\xi - a)^{\alpha-1} \tag{42}$$

Importantly, Equations (30) - (36) have been used to turn the fractional differential system from Equation (5) into an integer order one but with a fractional character. In addition, one can notice with Equations (39) - (42) that in the limit of $\alpha \to 1$, $m_1(\xi) = 1, m_2(\xi) = 0$ while $\varphi(\xi) = 1$, and Equation (39) converts to the case for the ordinary derivative. One can now handily solve for the potential, $\bar{v}_c$, and the current, $\bar{\iota}$, using conventional numerical techniques. For the present work, the system was solved by means of the fourth-order Runge-Kutta method, using a time step of $2\pi/1024\bar{\omega}$ [43].

## 3. Results

The one-parameter bifurcation diagram for the case of $\alpha = 1$ and $B_0 = 0.21$ was solved in a similar manner as that in [43] and is presented in Figure 3. These findings were found to agree with the results presented in the above work. Based on the figure, the fluctuations exhibit the chaotic behavior in the regions of $0 < B_1 < 2.8 \times 10^{-3}$ and $3.5 \times 10^{-2} < B_1 < 5.6 \times 10^{-2}$. For $5.6 \times 10^{-2} < B_1 < 6.4 \times 10^{-2}$, periodic doubling can be observed. In the other regions of the graph, there are only one-point attractors.

Figure 4 shows the potential, $\bar{v}_c(\xi)$, for $B_1 = 0.0065$ with $0 \leq \xi \leq 60$ and $0.6 \leq \alpha \leq 1$. For shorter times, the fluctuations are more dynamic in which the amplitude of the earlier oscillations decreases with respect to $\alpha$. Furthermore, the potential, $\bar{v}_c(\xi)$, attains a maximum and then subsequently exhibits the oscillation death. The onset of the decrease in the amplitude of the fluctuations appears to occur at earlier times as $\alpha$ is decreased.

The potential, $\bar{v}_c(\xi)$, for $B_1 = 0.0415$, where the oscillator exhibits the chaotic behavior for $\alpha = 1$ (see Figure 5), was plotted for $0 \leq \xi \leq 100$ and $0.6 < \alpha < 1$. For $\alpha = 1$, the chaotic behavior can be observed throughout the given interval, whereas the same behavior is only displayed for earlier times and for $\alpha$ equal to 0.6, 0.8, and 0.9. Furthermore, for the same $\alpha$ values, the oscillations appear to settle into the relaxation oscillations after the period of transient chaos. Interestingly, for $\alpha = 0.7$, the fluctuations exhibit the oscillation death.

The power, $P(t)$, into the capacitor is simply the product of the current and the voltage across the element. To write the power as a function of time, the normalized potential, current, and variable $\xi$, were converted back into their original form, as found in Equation (2). The above was done by setting the inductance L, γ, and β were set, respectively, to 1 Vs/A, 1 A/V, and 1 A/V³. Using these definitions, $P(t)$ can be written as:

$$P(t) = \begin{cases} C_\alpha v_c(t) {}_0 D_t^\alpha v_c(t) & 0 < \alpha < 1 \quad \text{(43a)} \\ C_\alpha v_c(t) i_c(t) & \alpha = 1 \quad \text{(43b)} \end{cases}$$

Using the relation $i_c(t) = \frac{d}{dt} v_c(t)$ (for $\alpha = 1$) and integrating the power gives the energy, $W(t)$, of the capacitor:

$$W(t) = \begin{cases} \dfrac{C_\alpha}{\Gamma(1-\alpha)} \int_0^t v_c(t') \left[ \int_0^{t'} \dfrac{\dot{v}_c(\tau)}{(t'-\tau)^\alpha} d\tau + \dfrac{v_c(0)}{t'^\alpha} \right] dt' & 0 < \alpha < 1 \quad \text{(44a)} \\ \dfrac{1}{2} C_\alpha v_c^2(t) & \alpha = 1 \quad \text{(44b)} \end{cases}$$

It should also be mentioned that for Equation (44a), the Caputo definition of the fractional derivative was used [55]. For our case, the second term, $\frac{v_c(0)}{t'^\alpha}$, vanishes due to the initial condition, $v_c(0) = 0$. However, if the potential is a constant non-zero value [$v_c(0) \neq 0$], the energy stored in this type of the capacitor will be proportional to $t^{1-\alpha}$. This result has been derived in the previous work involving fractional-order capacitors [56].

Based on Equation (44a), the energy stored in the non-ideal capacitor is equal to the integral of the applied voltage multiplied by a current that has inherent memory effects. Furthermore, setting $C_\alpha = \frac{\Gamma(1-\alpha)}{R^{\alpha-1} C^\alpha}$ for the coefficient from Equation (44a) and taking the limit as α → 0, one recovers the energy across a resistor. The above statement supports the idea that the non-ideal capacitor is a circuit element, which is transitory between a capacitor and a resistor.

To calculate the energy stored on the capacitor, the coefficient, $C_\alpha$, as proposed by Westerlund et al. [20], will be used:

$$C_\alpha = \begin{cases} \varepsilon_0 \varepsilon_r \omega^{1-\alpha} \sin\left(\frac{\pi\alpha}{2}\right)\frac{A}{d} & 0 < \alpha < 1 \\ \varepsilon_0 \varepsilon_r \frac{A}{d} & \alpha = 1 \end{cases} \quad \begin{matrix} \textbf{(45a)} \\ \\ \textbf{(45b)} \end{matrix}$$

Here $\omega$ is the angular frequency of the harmonically-varying field, $\varepsilon_0$ is the permittivity of the free space with a value of 8.854 x $10^{-12}$ As/Vm, $\varepsilon_r$ is the relative permittivity of the dielectric, A is the area of the capacitor plates, and d is the distance between them. Notice how Equation (45a) converts to Equation (45b) for $\alpha = 1$. To keep in line with the parameters used for this study, $\omega$ was chosen to be 1.35 Hz.

The values, $\varepsilon_r$, were found in the literature for various substances that corresponded to $\alpha$ values used in the present work. More specifically, the materials were polypropylene ($\alpha = 1$, $\varepsilon_r = 2.2$), stearic acid ($\alpha = 0.95$, $\varepsilon_r = 2.59$), anthracene ($\alpha = 0.85$, $\varepsilon_r = 3$), and silicone dioxide ($\alpha = 0.6$-$0.7$, $\varepsilon_r = 3.9$) [20, 21, 57-59]. To match the results of the current work, the fractional parameter was rounded down to 0.9, 0.8, and 0.6, for the stearic acid, anthracene, and silicone dioxide, respectively. Since $\varepsilon_r$ was not available for $\alpha = 0.7$, it was estimated by interpolating the values above, giving approximately 3.45 for the permittivity. From the listed values, it is evident that the fractional exponent decreases with respect to this parameter.

Figure 6 shows the energy stored on the capacitor as a function of $t$ for $0.6 \leq \alpha \leq 1$ with $B_1 = 0.0065$. Here the capacitor was assumed to have a volume of 1 cm$^3$ (d = 1 cm, A = 1 cm$^2$). As can be seen in the graph, the energy increases in a monotonic fashion with respect to $t$ for $0.6 \leq \alpha \leq 0.9$. In addition, the curves decrease with respect to $\alpha$.

Figure 7 shows the energy stored on the capacitor for $B_1 = 0.0415$. Unlike Figure 6 (for $B_1 = 0.0065$), the energy exhibited observable fluctuations for $\alpha < 1$. However, like Figure 6, $W(t)$ was larger for $\alpha < 1$. On the other hand, the energy associated with $\alpha = 0.7$, where there was the oscillation death, attained higher values for $t > \sim 4$ s. Like the potential values, the energy for the same $\alpha$ displayed the smallest oscillations, as compared to the other exponents.

Figure 8 presents the one-parameter bifurcation diagrams for $B_1$ ranging from 0 to 0.08 as a function of the fractional exponent, $\alpha$. It should also be stated that the points for $\alpha = 1$, at a given $B_1$ value, correspond to the $\bar{v}_c(\xi)$ values that can be observed in Figure 1. When $B_1 = 0$, the potential tends to reach a single value of $\sim 0.555$ until $\alpha = 0.99$, where the chaotic behavior emerges. For $B_1$ ranging from $0.01 - 0.03$, the curve does not exhibit the chaotic behavior for the range of fractional exponents. In addition, the fixed point generally increases until it reaches a maximum and then subsequently decreases. Furthermore, this peak also appears to be narrow, as $\alpha$ increases.

Once $B_1$ becomes larger than 0.03, however, a few things occur. For $B_1 = 0.04$, a bifurcation occurs at $\alpha \sim 0.955$, and there is also erratic behavior for $\alpha = 0.83$, which is followed by a drop in $\bar{v}_c(\xi)$ at $\alpha = 0.84$. Moreover, the region of the chaotic behavior near $\alpha = 0.83$, as exhibited for $B_1 = 0.04$, increases in size as the sinusoidal voltage increases. On the other hand, the region containing both the bifurcation point and chaotic behavior disappears for $B_1 = 0.07$.

To examine the interplay between the memory effects and chaotic behavior of the potential in the non-ideal capacitor, $\bar{v}_c(\xi)$ was plotted for $\xi$ values ranging from 150 to 1,400 with $B_1 = 0.06$ and $\alpha$ ranging from 0.6 to 0.9, as shown in Figure 9. To give a more detailed picture of the chaotic and non-chaotic behavior of the above figure, a magnification of the important regions is displayed in Figure 10. As can be seen in the figures, the time until the onset of chaos increases, as $\alpha$ is increased. Additionally, the fluctuations are quasiperiodic for $\alpha = 0.9$, for the given interval. Furthermore, the relaxation oscillations (for $\alpha \leq 0.7$) that occur after the chaotic behavior are smaller in magnitude, as compared to the fluctuations beforehand. It should also be stated that for $0.058 < B_1 \leq 0.08$ and $0 \leq \xi \leq 60$, the curves displayed the nearly-identical quasiperiodic behavior.

The corresponding phase planes for the results from Figure 8 can be observed in Figure 11. For each graph, the current, $\bar{\imath}(\xi)$, was plotted with respect to $\bar{v}_c(\xi)$. As can be seen, the curves for $\alpha < 0.9$ possess unstable limit cycles [14] that are indicative of the chaotic behavior. Moreover, the size of the attractor region is maximized for $\alpha = 0.7$, indicating that the oscillations are more irregular, as compared to the other conditions in the given domain. In contrast, the fluctuations for $\alpha = 0.9$ are quasiperiodic in which the behavior is signified by a phase portrait that contains a stable limit cycle.

To gain a better understanding of how the current moves down the ladder system, the ratios $R_k/R_0$ and $C_k/C_0$ elements were calculated, using Equations (23) and (24), and are displayed in Figures 12 and 13 for k = 1 to 20. As can be observed in the graphs, the values appear to increase or decrease in a monotonic fashion. Interestingly, the resistor values appear to behave in an opposite manner, as compared to the capacitor elements. It can be seen from Figure 12 that for $0.6 < \alpha < 0.9$, $R_k$ increases with respect to $k$, in which the magnitude of the curves range from ~2 to 200. Conversely, the curves for the capacitance, as shown in Figure 13, range from ~1 to ~0.2 where they decrease with increasing k. Furthermore, the curves for $R_k$ increase in value with respect to $\alpha$, while $C_k$ decrease with respect to this parameter.

## 4. Discussion

### 4.1 Fluctuation Modeling and Analysis

As observed in Figure 6, the amount of the stored energy in the capacitor increases with decreasing $\alpha$. Hartley et al. derived a similar relationship for the energy gained on a fractional order capacitor exposed to a constant charging voltage [56]. As discussed earlier, the fractional exponent, $\alpha$, corresponds to the degree of memory in the system, and can, thus, be thought of as a memory parameter. More specifically, as $\alpha$ tends towards 1, the scheme tends to reach a memoryless (Markovian) system [60, 61]. Thus, as the memory of the capacitor system increases, the energy stored in the dielectric increases over time, as well.

Westerlund suggested that the memory effect in a non-ideal capacitor is closely related to the dielectric absorption in the material [28]. He related this effect with the ability of the dielectric in the capacitor to "remember" previous voltages. This memory phenomenon was observed in experiments that involved an open capacitor that was previously charged and discharged for a predetermined set of time [20]. It was found that the potential curves increased in value when the fractional exponent corresponding to the dielectric decreased in value.

Further insight into the above phenomena can be gained by examining the types of dielectric materials used in the same study. The materials reported include polypropylene and polyvinylidene fluoride with fractional exponents of 0.999952 and 0.9776, respectively. As discussed previously, the decrease in the fractional exponent was linked to an increase in the relative permittivity, $\varepsilon_r$, of the dielectric. For this study, it was 2.2 for polypropylene and 11 for polyvinylidene, which indicates that the memory and the stored energy in the capacitor are linked to the polarizability of the constituent elements in the dielectric, and, hence, the microscopic makeup of the material.

A more explicit link between the energy stored on the capacitor and the memory of the system can be observed in the central integral of Equation (44a). Here, the integrand involves the product of the time derivative of the potential and a power-law memory kernel with $\alpha$ as the exponent. Additionally, the potential is proportional to the applied electric field on the capacitor. Therefore, it can be thought that the amount of the electrostatic energy stored on the capacitor increases as its sensitivity to the past dynamical behavior of the field increases. In other words, an increased ability of the dielectric to "remember" past excitations result in a larger amount of the energy contained therein. However, the underlying causes of this behavior are not well understood.

This interplay between the memory and the energy stored on the capacitor system can also be observed in Figures 5 and 7. As can be observed in the graphs, the stored energy is the highest for $\alpha = 0.7$, where the oscillation death occurred in which the voltage values fluctuate the least. It is apparent that the most energy is stored when the oscillations are minimized. Moreover, the effects of memory on the stored energy can be seen in Equation (44a), where the reduction in the energy arises from the integral taking account of the negative change in the voltage with respect to time.

The implications of the behavior of the resistor and capacitor elements, as observed in Figures 12 and 13, may be explained as follows. As $\alpha$ is decreased, the curves of the $R_k$'s decrease while those for the $C_k$'s increase. This relationship not only signifies an increase in the ability of the current to travel down the ladder but the degree to which the corresponding energy can be stored in each subsequent capacitor element. Since a decrease in $\alpha$ signifies an increase in the memory of the system, the capacitor may be thought of as a memory element in addition to a unit of the energy storage.

When $\alpha \to 0$, the $R_k$'s become negligible while the $C_k$'s become arbitrarily large. This behavior indicates that the current propagates down the ladder unimpeded, and the voltage on the capacitor element becomes vanishingly small due to the relation of $C = q/V$, where $q$ is the charge. Furthermore, Ohm's law is recovered for the overall system where it behaves like a resistor. In this condition, the term, $C_0$, represents the parasitic capacitance in the element [62].

On the other hand, as the fractional parameter trends towards 1, the resistance in each consecutive element grows at a fast rate with increasing $k$. In addition, the capacitance on each element, $C_k$, decreases in a similar fashion. In this sense, perhaps, a resistor element may be thought of as a memory negator through some types of dissipation mechanisms. As can be observed in Equations (23) and (24), in the limit of $\alpha \to 1$, all $R_k$ tend to infinity while all $C_k$ become infinitesimally small. This trend symbolizes the inability of the current to traverse down the ladder and store energy there, and thus the ladder becomes a simple capacitor where $R_0$ represents the equivalent series resistance.

When $B_1 = 0.06$, a decrease in $\alpha$, and consequently an increase in the memory of the system, leads to an earlier onset of chaos, as observed in Figure 9. The memory-induced chaos has also been observed in the cardiac excitation [63]. For the given parameters, therefore, a reduction in the memory of the system can delay the onset of dynamical instabilities in the potential across the capacitor. Perhaps the ability of the dielectric to remember earlier values may lead to greater sensitivity in initial conditions, which consequently induces chaos at earlier times. The reason for the greater degree of the erratic behavior for $\alpha = 0.7$, as compared to the other conditions (see Figures 10 and 11) is not fully understood at this time, and, therefore, future work will involve resolving this issue.

## 4.2 Thermal Voltage Noise Analysis

The power density spectrum of the equilibrium voltage fluctuation on an impedance, $Z(\omega)$, according to the fluctuation-dissipation theorem, is given by [64]:

$$S_V(\omega) = 4k_B T Re\{Z(\omega)\} \tag{46}$$

where $k_B$ is Boltzmann's constant, $T$ is the temperature, $h$ is the Planck's constant, and $\omega$ is the angular frequency of the oscillations. Using the universal dynamical theory, Teitler et al. [65] was able to derive the following spectral dependence:

$$S_V(\omega) \propto \frac{1}{\omega^\alpha} \quad ; \ 0 < \alpha < 1 \ ; \ \omega \gg \omega_p \tag{47}$$

Since the exponent from Equation (47) is the same as that used for the Curie-von Schweidler law, it is also equal to the exponent of the fractional derivative in the context of the present work. Therefore, it can also be said that the spectral dependence depends on the memory of the system. In the extreme cases, the voltage exhibits white noise fluctuations for $\alpha = 0$, while displaying the pink (flicker) noise behavior for $\alpha = 1$. In terms of the capacitor system, the pink noise is, thus, associated with a dielectric that is memoryless, while the white noise corresponds to one with the complete memory. Since this type of noise is associated with direct current [67], it will be assumed that this kind of behavior will only apply to the system when $V_1 = 0$ and the current has reached the steady-state condition.

Ngai et al. suggested that the spectral exponent is affected by the microscopic configuration of the dielectric [65]. For example, the exponent may change if the material has become more heterogenous over time due to an applied stress. Howell et al. found that $\alpha$ varied as a function of temperature in a study involving the dielectric relaxation of a glass forming material [72]. They hypothesized that the change in the decay behavior (varying $\alpha$) arose from an increase in the microscopic heterogeneity with increasing temperature at constant pressure.

Brechtl et al. previously examined the complexity of the $\frac{1}{\omega^\alpha}$ noise (also referred to as the colored noise) with exponents ranging from -2 to 2 using the refined composite multiscale entropy technique [68, 69]. As discussed in [70-72], the complexity is a measure of the degree of long-range correlations and intricacy of the dynamical behavior of a given system. To model and analyze the time series, using this method, the complexity metric is plotted with respect to the scale factor, which is used to coarse-grain the original time series. The complexity at a given scale factor is typically called the sample entropy value [69]. For series in which the sample entropy either increases or remains constant with an increasing scale factor, the series consists of long-range correlations.

Figure 14 displays the sample entropy for the colored noise for $\alpha$ values ranging from 0 (white noise) to 1 (pink noise). As can be seen, the curves increase, overall, with an increase in the above parameter. Additionally, the sample-entropy goes from a strictly-decreasing function at $\alpha = 0$ to remaining practically constant at $\alpha = 1$. What is also evident from the graph is that there is a spectrum of sample-entropy curves with respect to the exponent, $\alpha$. Using the above ideas, it can, therefore, be said that the degree of long-range correlations in the colored noise increases in a continuous fashion as the spectral exponent rises in value.

Now we will examine the behavior of a capacitor which contains a liquid dielectric. In this context, the dielectric will be composed of charge carriers instead of dipolar elements, where $\alpha$ ranges from 0.6-0.95 [31]. For this system, the motion of the carriers will governed by the Langevin equation subject to a stochastic (colored noise) force. This force will result from the bombardment of the carrier by the molecules or ions of the electrolyte [67].

Monnai et al. examined the effect of the spatio-temporal colored noise on the diffusive nature of a particle in a potential well [73]. It was determined that an increase in the spatial correlations of the noise corresponded to a decrease in the drift velocity of the particle. Conversely, it can be thought that a decrease in the spatial correlations leads to an increase in the same velocity.

Based on the discussion above, it can be thought that the degree of long-range correlations of the voltage noise decreases with respect to the fractional parameter. Assuming these types of correlations have a spatial component, it will be suggested here that the drift velocity of the charge carriers will increase as $\alpha$ decreases. Since the drift velocity is proportional to the conductance of the media, the above hypothesis suggests that the dielectric becomes more "metal-like" as the exponent decreases [28]. With respect to the ladder model, the decrease in the resistances, $R_k$, with decreasing $\alpha$, can be thought of as an analogue for this type of behavior.

It should also be mentioned that in [73], they determined that the spatial correlations were positively correlated with the Gaussian memory. The above finding is in contrast with the present work, where a higher amount of complexity in the voltage fluctuations was associated with a lower degree of memory. The link between the increase in the memory of the non-ideal capacitor and the decrease in the complexity of voltage fluctuations is not entirely understood and, therefore, should be explored in the future.

The decrease in the complexity of the noise may be due to an increase in the degree of inherent randomness in the system [70]. This increased randomness may be linked to the increased drift velocity of the carriers, where the behavior is assumed to be more erratic in nature. Thus, it may be said that an increasing metallic character corresponds to a decrease in the complexity of the noise fluctuations. In addition, if it is assumed that the level of impurities in the material remains constant, then an increase in the mobility leads to a rise in the trapping length [74]. Consequently, the interactions between the carriers and the impurities is decreased. A decrease in the amount of possible interactions in a dynamical system has been linked to the reduction in the complexity of the system behavior [75].

## 5. Conclusions

For the present work, the potential fluctuations of the Bonhoeffer van der pol oscillator system with a non-ideal capacitor was investigated, using a fractional differential equation model. It was found that the order of the fractional derivative, $\alpha$, is also a measure of the memory in the dielectric. Furthermore, it was found that the dynamical behavior of the potential across the capacitor was affected by this parameter, and, therefore, the memory of the system.

Importantly, findings indicate that an increase in the memory was associated with an increase in the energy stored in the dielectric. Therefore, it can be theorized that the fractional derivative may also be some type of function of the stored energy, although the underlying reasons for this trend are not fully understood. Furthermore, the effects of the dynamical behavior on the capacity of the dielectric to store energy was observed. Here, fluctuations that exhibited the oscillation death led to more energy being stored in the dielectric, as compared to the behavior, which displayed relaxation oscillations and chaotic behavior. The relatively-lower stored energy resulting from the latter types of oscillations is accounted for by the memory effect, where the present accumulation of energy is affected by previous decreases in the potential across the component. Thus, the dielectric material "remembers" the past behavior of the voltage, which result in either a decrease or an enhancement in the stored energy.

In the state in which $B_1 = 0.0065$, where the system displayed no chaotic behavior, the oscillations were seen to decrease in magnitude with a decrease in the fractional parameter. For the case where $0.6 \leq \alpha \leq 0.9$, the energy was found to increase in a monotonic fashion with respect to the normalized time, $\xi$, and the overall curves increased with respect to $\alpha$. The same type of behavior was exhibited in the chaotic state of $B_1 = 0.0415$, except at $\alpha = 0.7$ (the oscillation death), it displayed a similar trend with respect to both parameters. In addition, the effect of the history of the voltages across the capacitor was apparent where the fluctuation exhibiting the oscillation death led to the highest amount of energy stored on the capacitor, as compared to the other conditions after enough time had elapse.

Moreover, an increase in $\alpha$, under certain conditions, led to the earlier onset of chaotic voltage oscillations across the capacitor.  It is suggested here that perhaps an increase in the memory of the system causes the behavior to be more sensitive to its initial state, which results in erratic fluctuations at earlier times.  Additionally, the corresponding phase portraits showed that the chaotic behavior was enhanced, in general, with a decrease in $\alpha$.

In addition, it was found that the non-ideal capacitor has a transitory nature, where it becomes more like a resistor as $\alpha \to 0$, and conversely, more like a capacitor as $\alpha \to 1$. Furthermore, a decrease in $\alpha$ was linked to an enhanced metallic character of the dielectric due to the charge-carrier analogy.  This change was thought to be caused by an increase in the drift velocity due to the decrease in the long-range correlations of the thermal voltage noise.

## *Acknowledgements*

The authors would like to thank Dr. Shakoor Pooseh for helpful discussions concerning the implementation of the fractional differential equation algorithm.

*Figures*

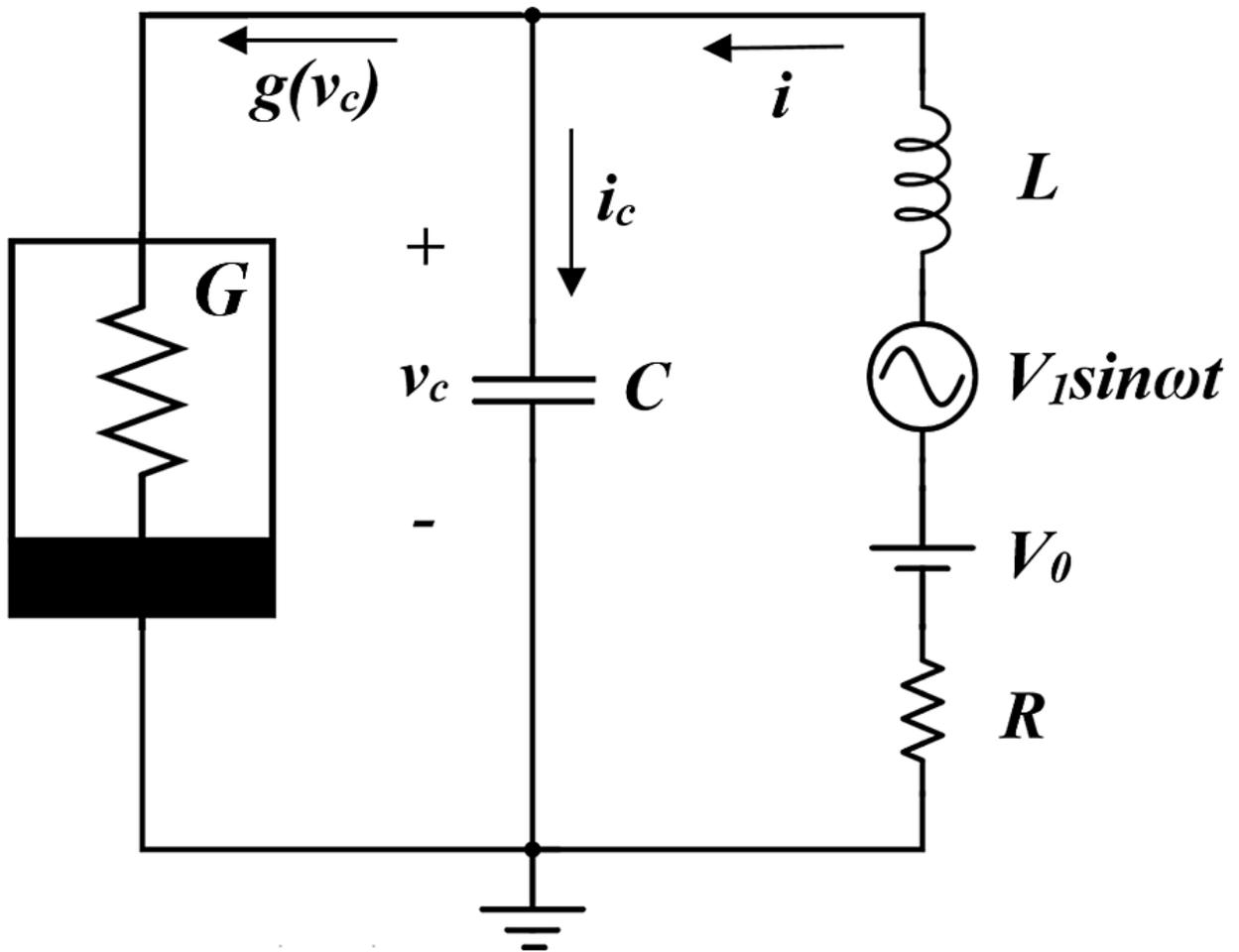

Figure 1. The circuit diagram for the Bonhoeffer Van der Pol Oscillator system.

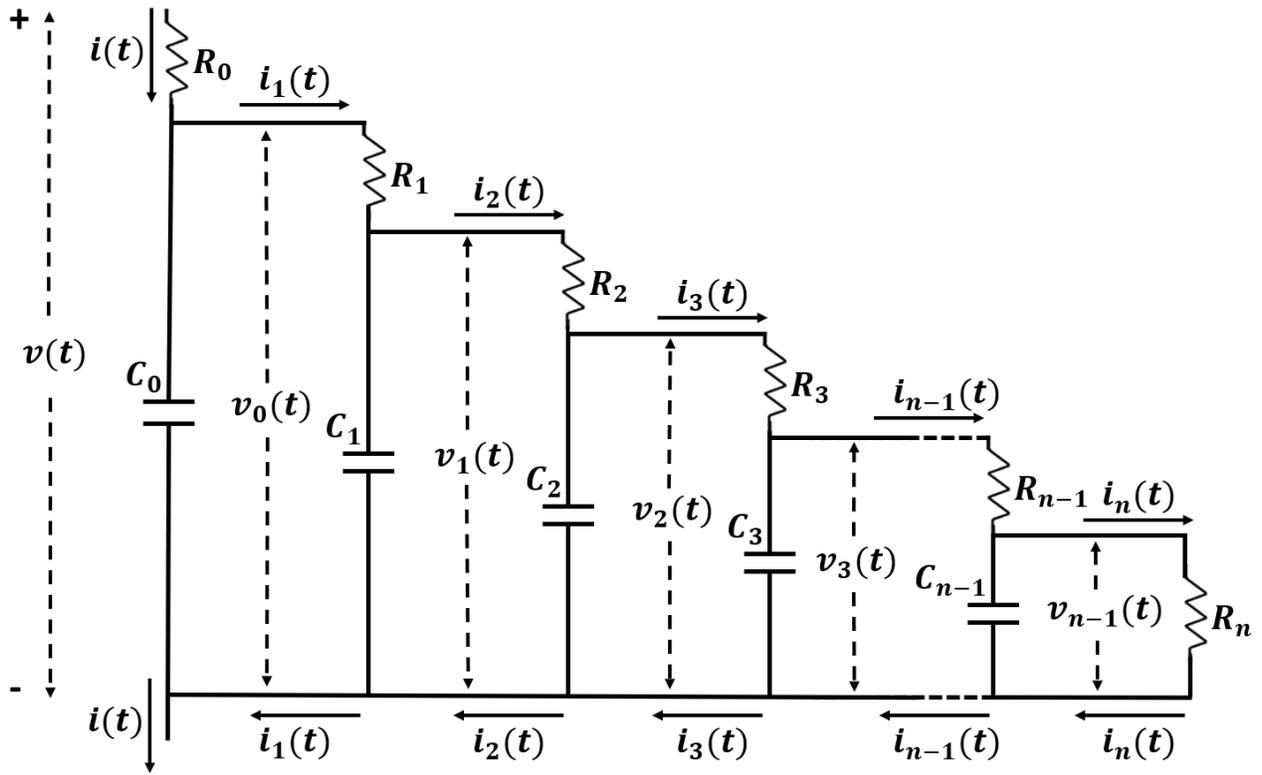

Figure 2. The RC hierarchical ladder model.

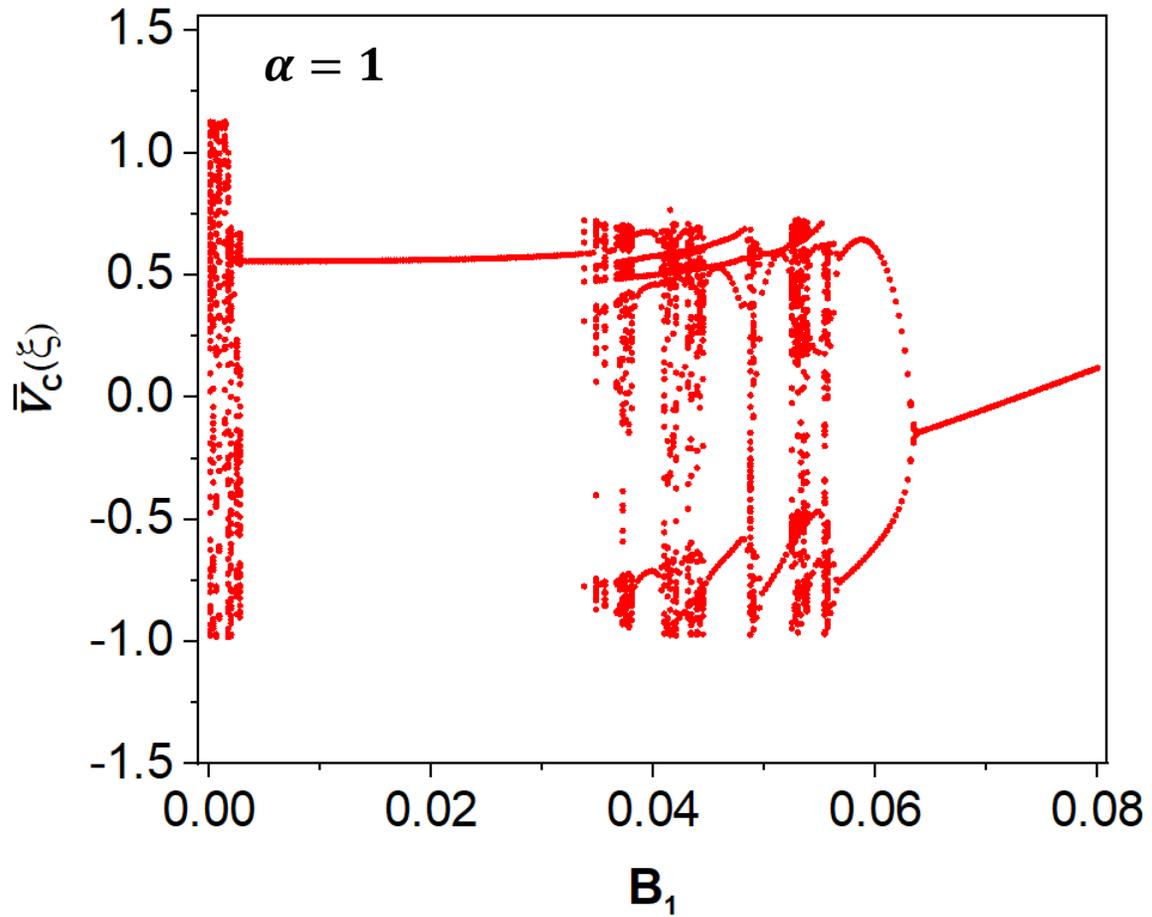

Figure 3. The one-parameter bifurcation diagram for the potential oscillations for the case of $\alpha = 1$.

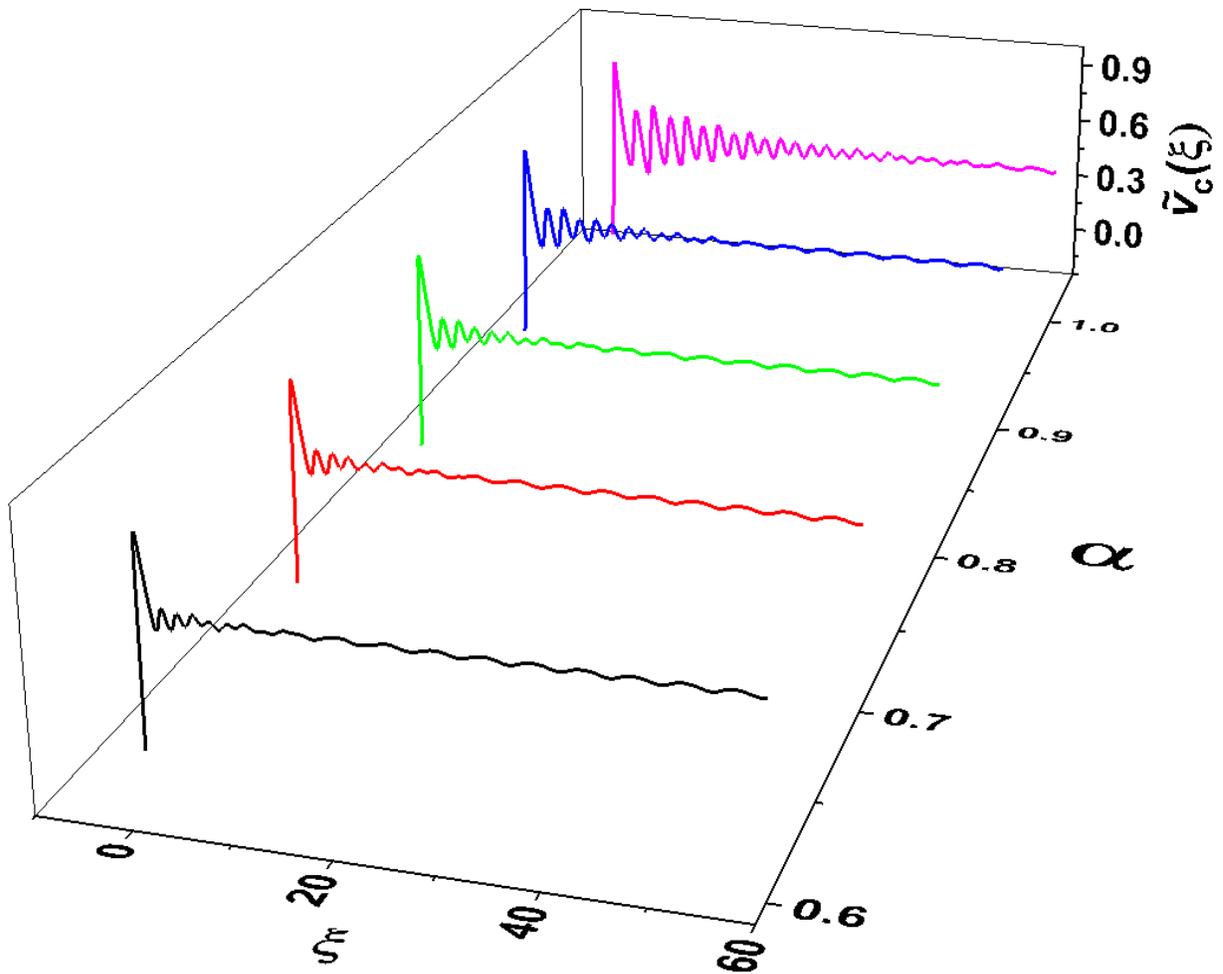

Figure 4. The potential $\bar{v}_c(\xi)$ for $0 < \xi < 60$, $0.6 < \alpha < 1$, and $B_1 = 0.0065$.

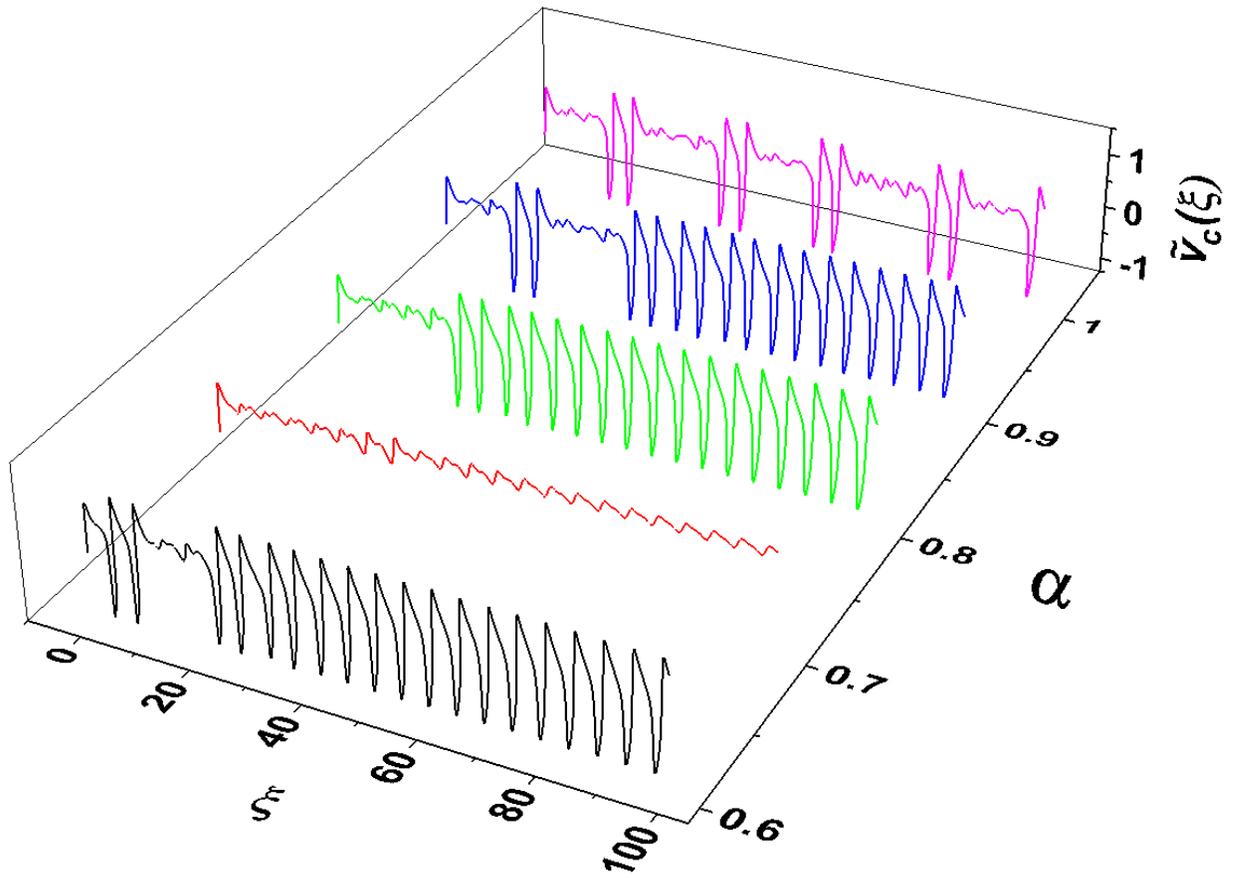

Figure 5. The potential, $\bar{v}_c(\xi)$, for $0 < \xi < 100$, $0.6 < \alpha < 1$, and $B_1 = 0.0415$.

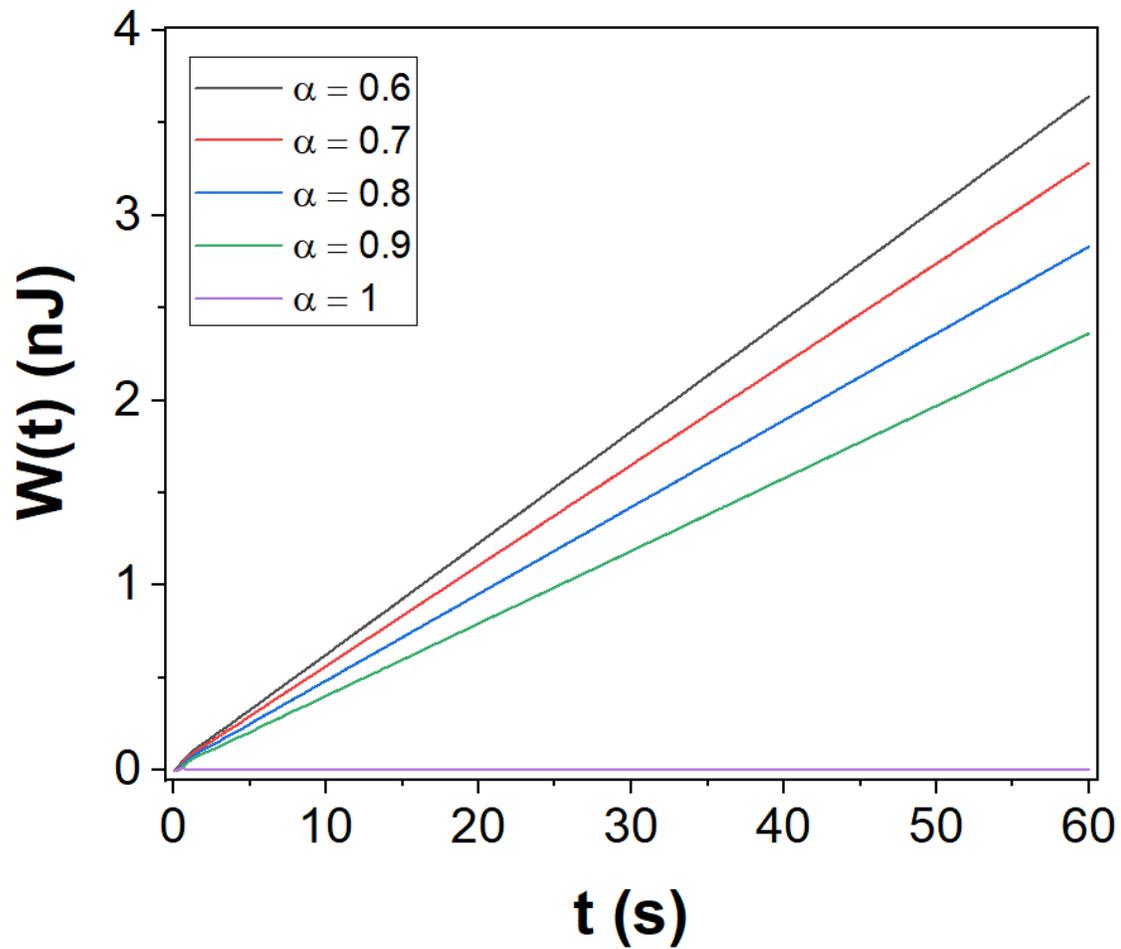

Figure 6. The energy stored on the capacitor, $W(t)$, for $0 \leq t \leq 60$, $0.6 < \alpha < 1$, and $B_1 = 0.0065$.

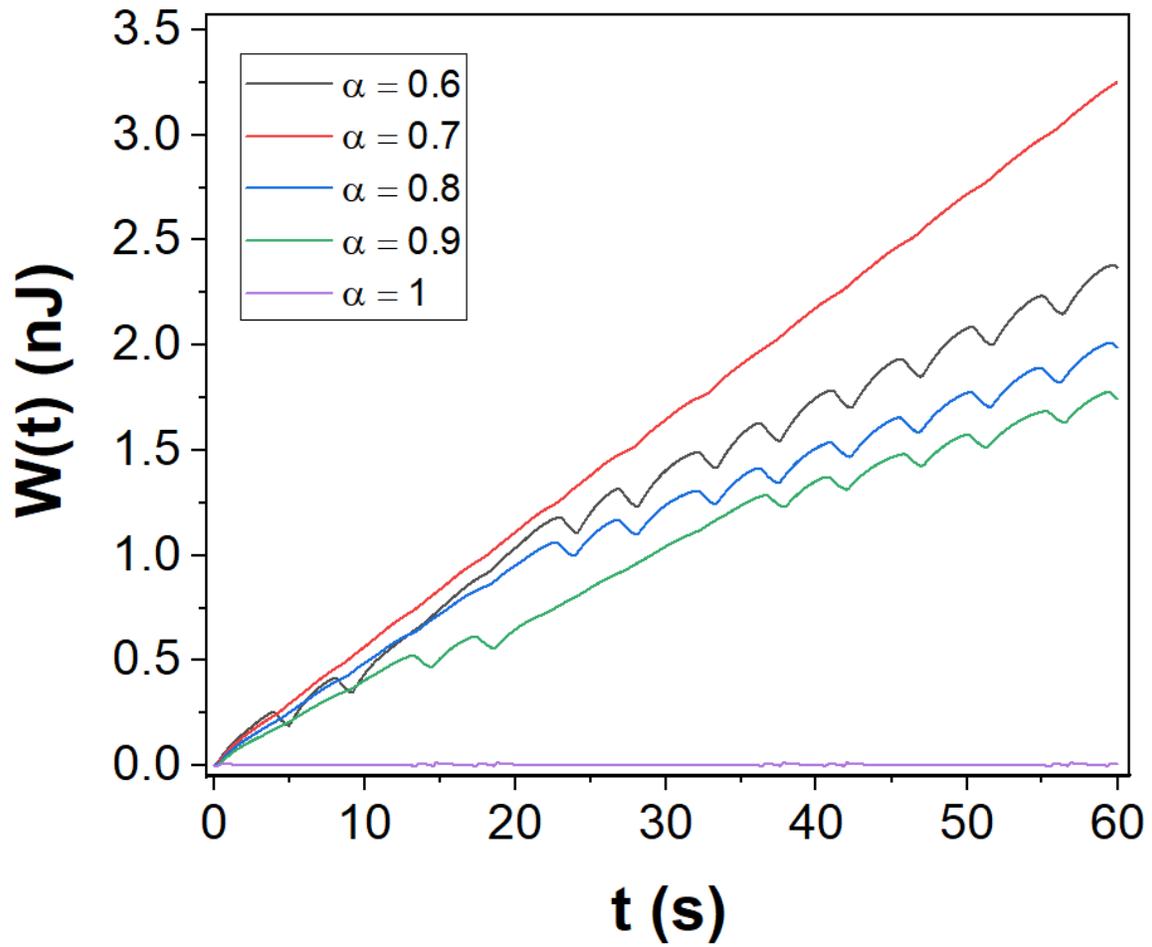

Figure 7. The energy stored on the capacitor, $W(t)$, for $0 \leq t \leq 60$, $0.6 < \alpha < 1$, and $B_1 = 0.0415$.

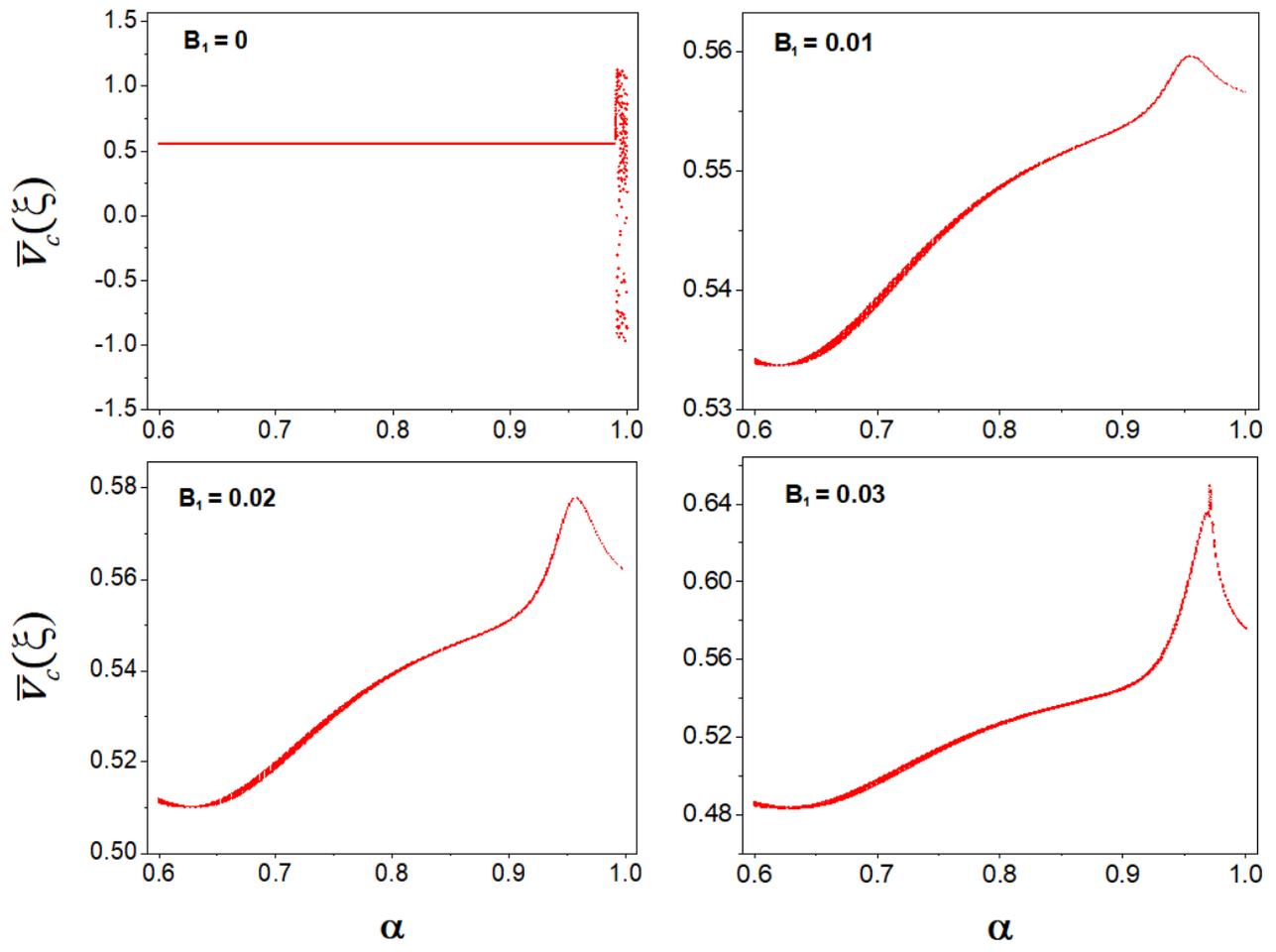

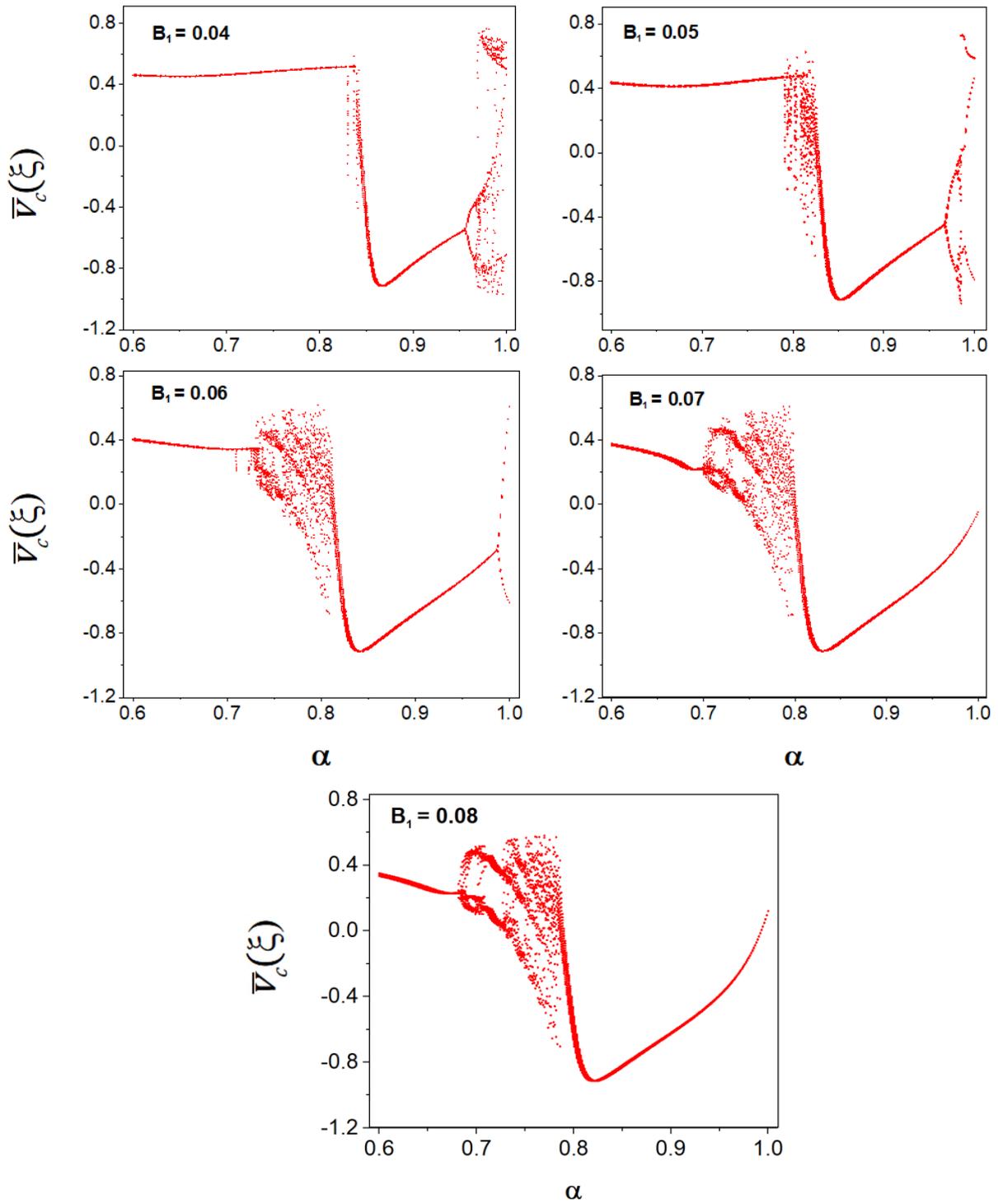

Figure 8. The one-parameter bifurcation diagram for the potential oscillations corresponding to $0.6 < \alpha < 1$ and $B_1$ ranging from 0 to 0.08.

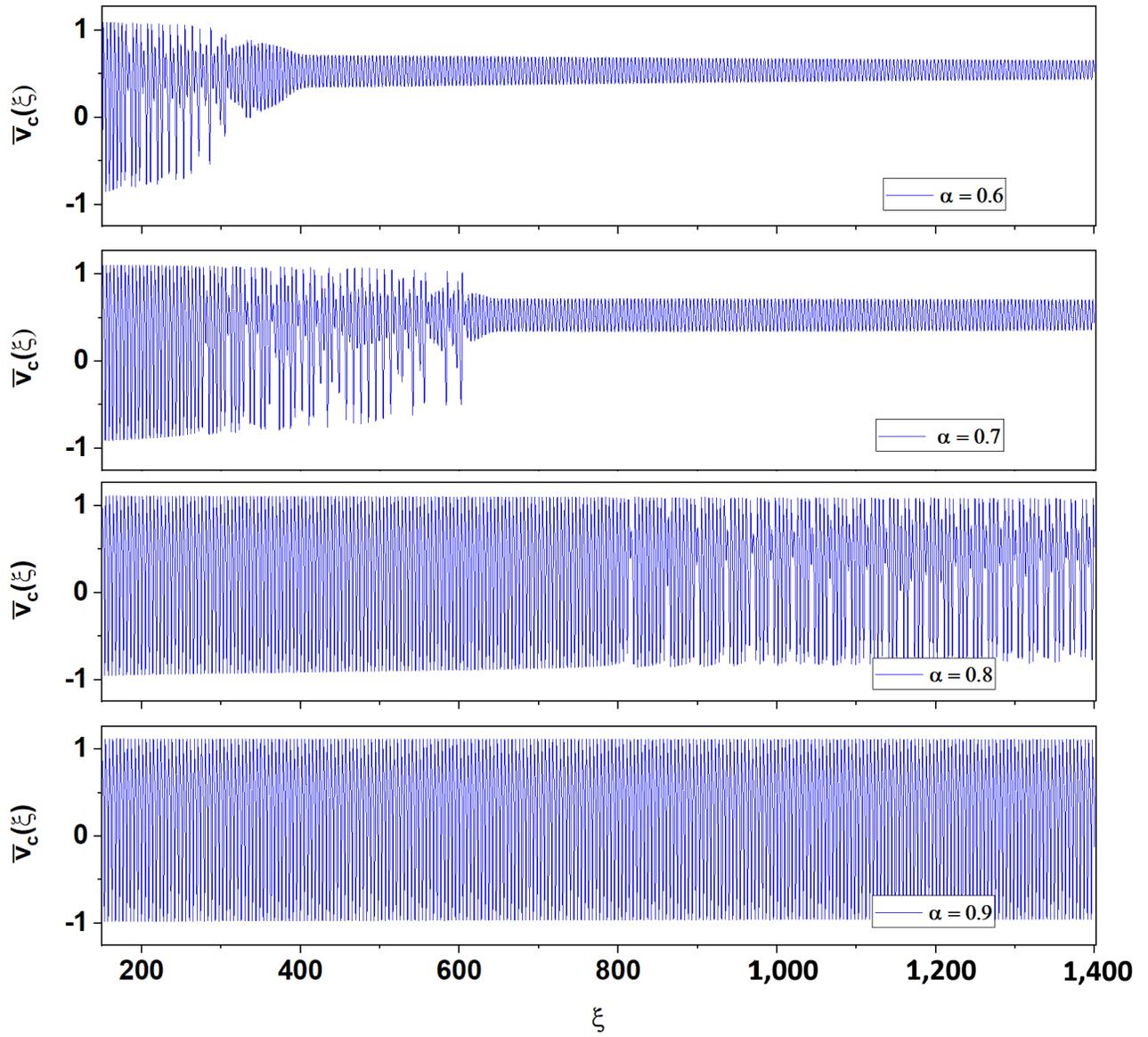

Figure 9. The potential, $\bar{v}_c(\xi)$, for $150 \leq \xi \leq 1,400$, $0.6 < \alpha < 0.9$, and $B_1 = 0.06$.

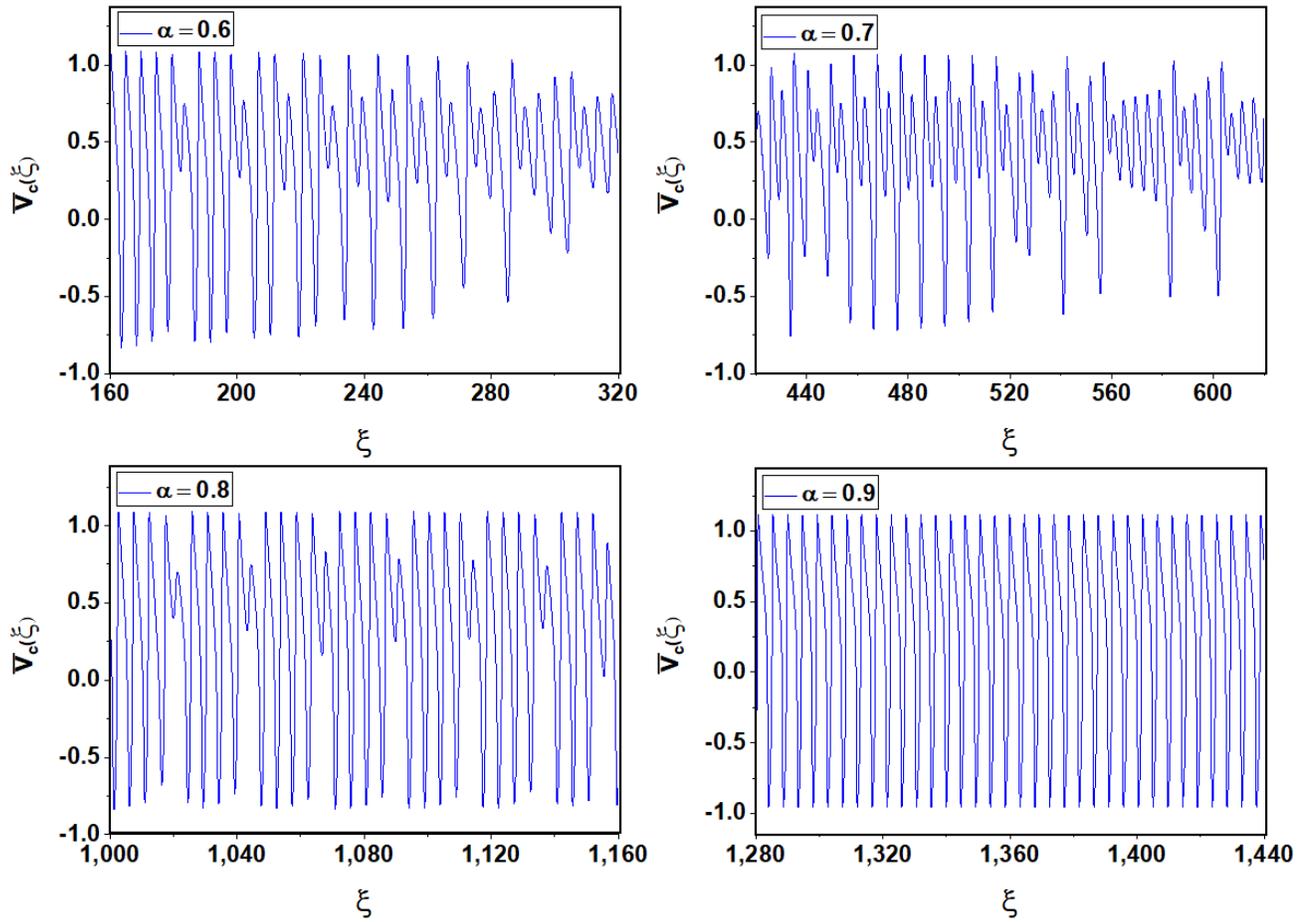

Figure 10. Magnified regions of chaotic behavior (0.6 < α < 0.8) and non-chaotic behavior for α = 0.9 from Figure 9 for $B_1 = 0.06$.

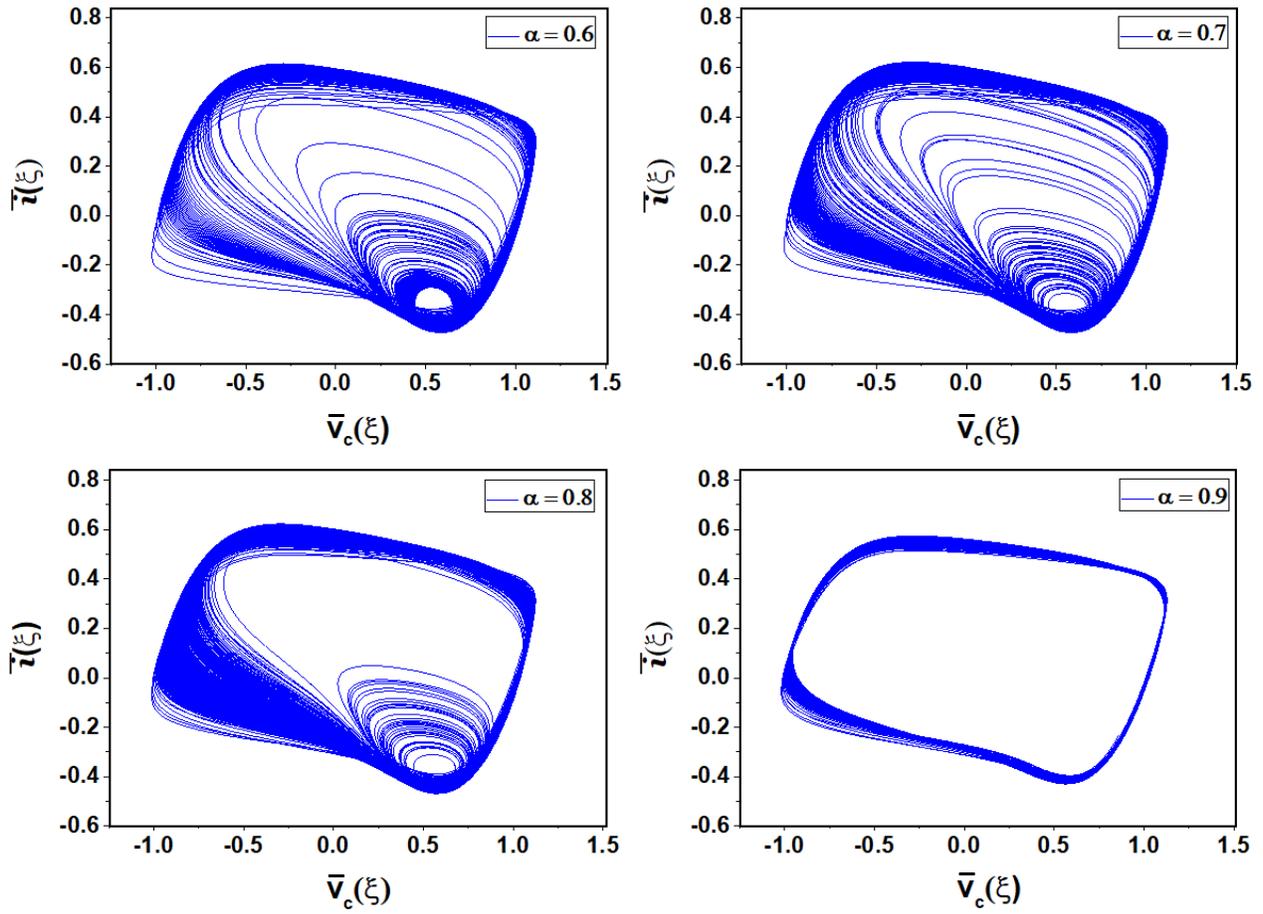

Figure 11. Corresponding phase portraits for Figure 9 (0.6 < α < 0.9) and non-chaotic behavior from Figure 9 for $B_1 = 0.06$.

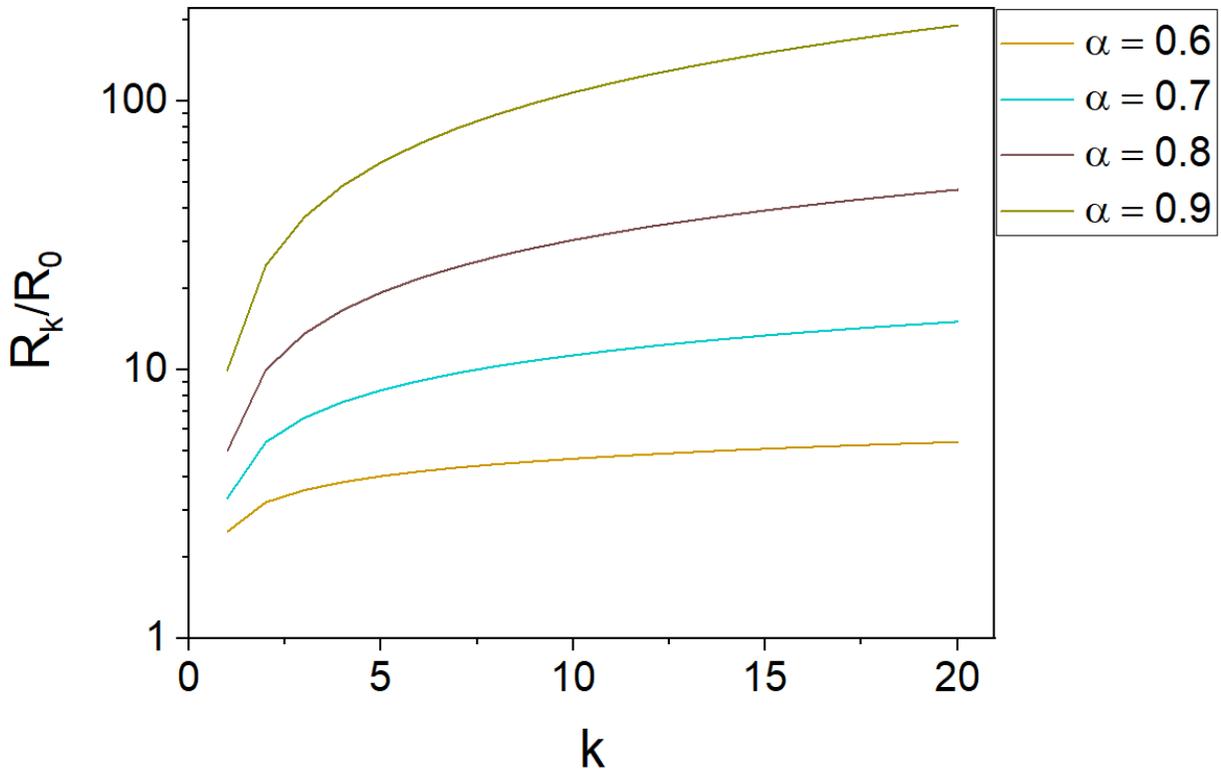

Figure 12. Ratio of the kth resistor element to the initial resistor element, $R_0$, from the ladder model plotted with respect to k for $0.6 < \alpha < 0.9$.

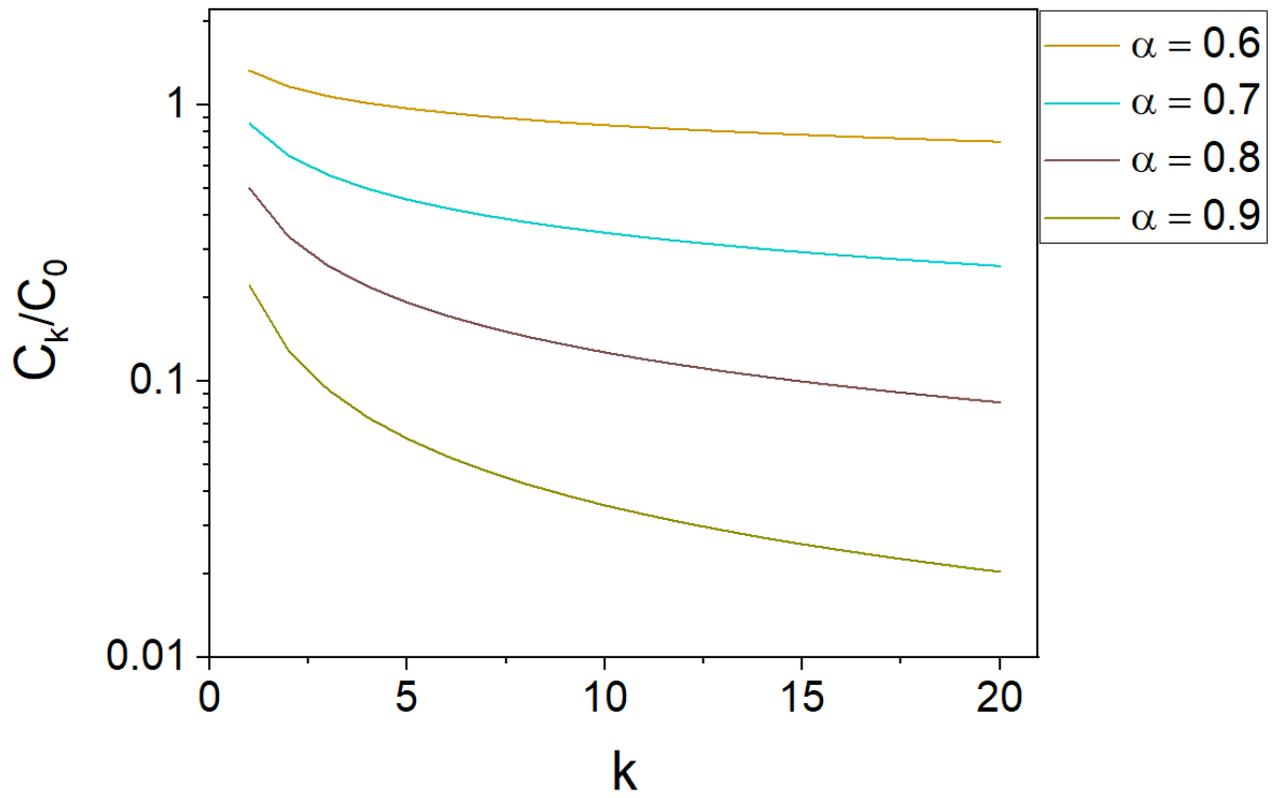

Figure 13. Ratio of the kth capacitor element to the initial capacitor element, $C_0$, from the ladder model plotted with respect to k for $0.6 < \alpha < 0.9$.

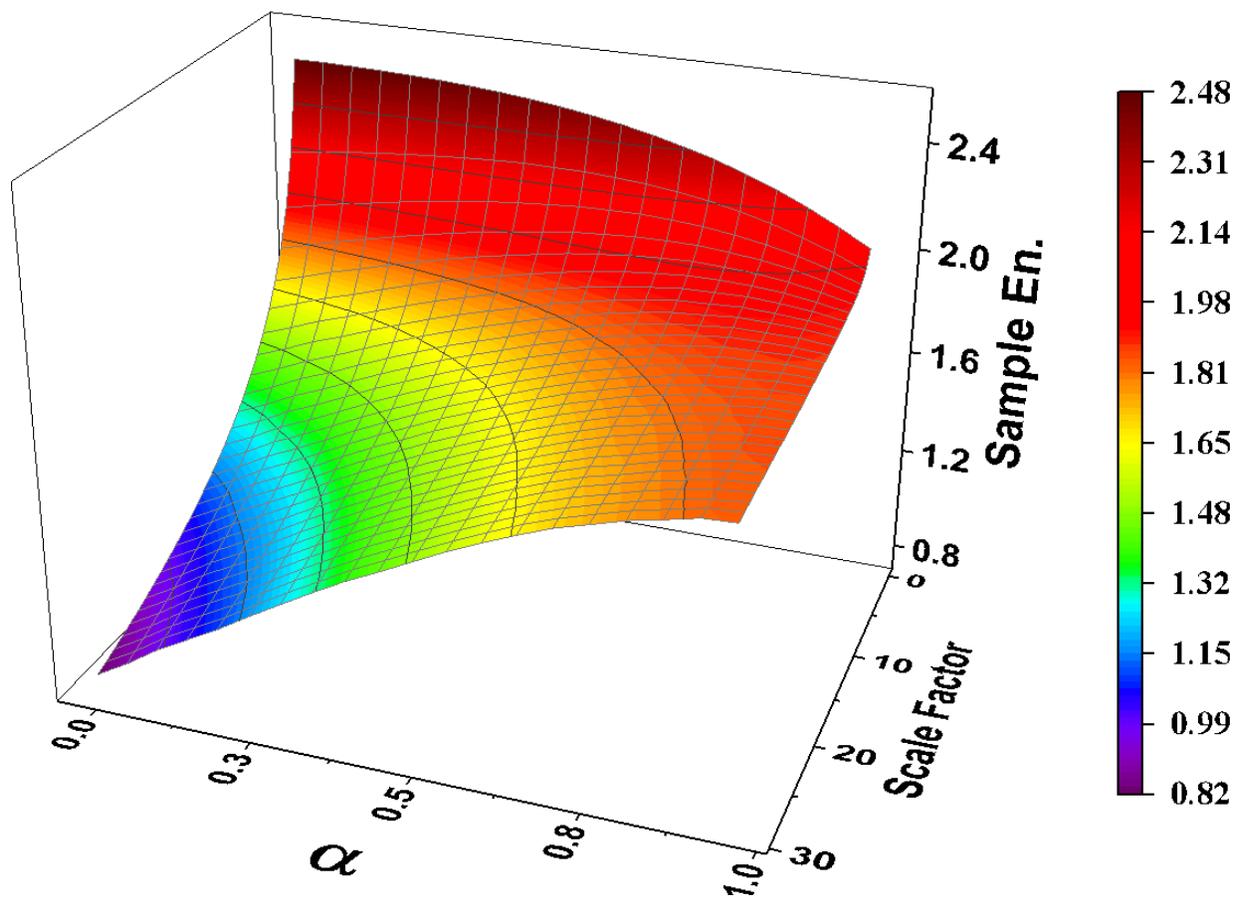

Figure 14: Mean sample entropy of the colored noise data for $1 \leq \tau \leq 30$ plotted for $0 < \alpha < 1$ (from reference [67]).